

\input mn
\input epsf


\let\sec=\section
\let\ssec=\subsection


\font\japit = cmti10 at 11truept
\def\ss{\scriptscriptstyle\rm}
\def\japref{\parskip =0pt\par\noindent\hangindent\parindent
    \parskip =2ex plus .5ex minus .1ex}
\def\gs{\mathrel{\lower0.6ex\hbox{$\buildrel {\textstyle >}
 \over {\scriptstyle \sim}$}}}
\def\ls{\mathrel{\lower0.6ex\hbox{$\buildrel {\textstyle <}
 \over {\scriptstyle \sim}$}}}
\newcount\equationo
\equationo = 0

\def\leftdisplay#1$${\leftline{$\displaystyle{#1}$
  \global\advance\equationo by1\hfill (\the\equationo )}$$}
\everydisplay{\leftdisplay}

\def\japfig#1#2#3#4{
\ifnum #2 = 1
\beginfigure{#1}
\epsfxsize=8.2cm
\centerline{\epsfbox{#3}}
\fi
\ifnum #2 = 2
\beginfigure*{#1}
\epsfxsize=15.5cm
\centerline{\epsfbox{#3}}
\fi
\ifnum #2 = 3
\beginfigure{#1}
\epsfxsize=8.2cm
\centerline{\epsfbox[60 208 510 588]{#3}}
\fi
\ifnum #2 = 4
\beginfigure{#1}
\epsfxsize=8.2cm
\centerline{\epsfbox[53 15 465 785]{#3}}
\fi
\ifnum #2 = 5
\beginfigure{#1}
\epsfxsize=6.5cm
\centerline{\epsfbox{#3}}
\fi
\caption{%
{\bf Figure #1.}
#4
}
\endfigure
}



%

\pageoffset{-0.8cm}{0.2cm}




\begintopmatter  

\vglue-2.2truecm
\centerline{\japit To appear in Monthly Notices of the R.A.S.}
\vglue 1.7truecm

\title{Testing anthropic predictions for ${\bf\Lambda}$ and the CMB temperature}

\author{J.A. Peacock}

\affiliation{%
Institute for Astronomy, University of Edinburgh,
Royal Observatory, Blackford Hill, Edinburgh EH9 3HJ}

\shortauthor{J.A. Peacock}

\shorttitle{Testing anthropic predictions for $\Lambda$ and the CMB temperature}


\abstract{%
It has been claimed that the observed magnitude of the vacuum energy
density is consistent with the distribution predicted in anthropic
models, in which an ensemble of universes is assumed. This calculation
is revisited, without making the assumption that the CMB temperature is known,
and considering in detail the possibility of a recollapsing universe.
New accurate approximations for the growth of perturbations and the
mass function of dark haloes are presented.
Structure forms readily in the recollapsing phase of a model with negative $\Lambda$,
so collapse fraction alone cannot forbid $\Lambda$ from being
large and negative. A negative $\Lambda$ is disfavoured only if we
assume that formation of observers can be neglected once the recollapsing
universe has heated to $T \gs 8$~K.
For the case of positive $\Lambda$, however, the current universe does occupy a
extremely typical position compared to the predicted distribution on the
$\Lambda-T$ plane. Contrasting conclusions can be reached if anthropic
arguments are applied to the curvature of the universe, and we discuss the
falsifiability of this mode of anthropic reasoning.
}


\maketitle  

\sec{INTRODUCTION}

There is now almost unanimous agreement that the universe contains
a component that strongly resembles Einstein's cosmological constant.
This conclusion is cross-buttressed by a variety of lines
of evidence (e.g. Spergel et al. 2006), and seems unlikely to change.
Physically, this means that the vacuum appears to possess a homogeneous
energy density, with an associated negative pressure.

The existence of a non-zero vacuum density raises two problems:
(1) the scale problem, and (2) the why-now problem. The
first of these concerns the energy scale corresponding to the
vacuum density. If we adopt the values $\Omega_v=0.75$ and $h=0.73$ for the
key cosmological parameters, then
$$
\rho_v = 7.51 \times 10^{-27}\, {\rm kg\, m^{-3}}
= {\hbar\over c}\, \left({E_v\over \hbar c}\right)^4,
$$
where $E_v=2.39$~meV is known to a tolerance of about 1\%.
The vacuum density should receive contributions of this form from the
zero-point fluctuations of all quantum fields, and one would expect a net
value for $E_v$ of order the scale at which new
physics truncates the contributions of
high-energy virtual particles: anything from 100~GeV to $10^{19}$~GeV.
The why-now problem further asks why we are observing the universe
at almost exactly the unique time when this strangely small
vacuum density first comes to dominate the cosmic density.

Taking the second problem first, a question that involves the
existence of observers must necessarily have an answer in which
observers play a role. Therefore, a solution to the why-now problem
requires anthropic reasoning. In defining exactly what this term
means, it is convenient to draw a distinction between what might be
called one-universe and many-universe anthropic arguments.  The first
of these should not be controversial, since it states that observers
are likely to exist at special times in this universe: complex
structures cannot begin to form until temperatures reach $T\ll 1000$~K,
and the formation of structure will switch off in
the near future when conditions become heavily vacuum dominated and
the universe enters a phase of exponential expansion. Because the
densities in baryons and dark matter are very roughly comparable,
and because the photon-to-baryon ratio is roughly $137^2 m_p/m_e$,
this window for structure formation also in practice
opens soon after matter-radiation equality (e.g. p97 of Peacock 1999),
although there is no suggestion that this is anything other than
a genuine coincidence. If we further assume that complex life
requires metals, it is no surprise
that the universe now has an age comparable to a typical stellar
lifetime, and this is almost certainly the explanation for Dirac's
large-number coincidence (e.g. Carter 1974).

The single-universe expectation that we should live relatively soon
after matter-radiation equality may be all that is required of anthropic
reasoning. The quintessence programme aims to find a dynamical
origin for the vacuum energy; the hope is that the change in cosmological
expansion history at matter-radiation equality may prompt the effective
vacuum density generated by some scalar field
to change from a sub-dominant contribution at early
times, to something that resembles $\Lambda$ by the present. This
would be a satisfying solution to the why-now problem, but it is not
clear that the mechanism can be made to work. For a simple scalar
field with an arbitrary potential, it seems that
an energy scale of $\sim 1$~meV needs to be introduced into the potential
by hand, in order to prevent the quintessence density from tracking
the overall mass density at all times (e.g. Liddle \& Scherrer 1999).
This simply swaps one unexplained coincidence for another.
Greater success is achieved with the more radical option of $k$-essence, in which
the field is given a non-canonical kinetic term. This readily achieves
departures from tracking, but no simple model has been exhibited in which
the late-time behaviour necessarily approaches something close to a
constant-density $\Lambda$ term (Malquarti, Copeland \& Liddle 2003).

The other major problem with the quintessence approach is that the
models do not solve the scale problem: the potentials asymptote to
zero, even though there is no known symmetry that requires this.
This leads to consideration of the more radical many-universe
mode of anthropic reasoning. Here, one envisages making many copies
of the universe, allowing the value of the vacuum density to vary
between different versions. The simplest concrete way of generating
this multiverse is via stochastic inflation with an additional
scalar field that sets the effective value of $\Lambda$
(e.g. Garriga \& Vilenkin 2000). In this paper we will not
need to be specific about the mechanism involved, although it
is of course of the greatest interest. But first we must establish
whether an ensemble approach makes sense. The logic is the same
as in evolutionary biology, where we start with empirical
evidence for selection out of a diversity of heredity, and only later move on to
the search for the microscopic mechanism of genes and DNA
that permits this diversity.

Although most members of the hypothetical
ensemble are presumed to have large vacuum densities
comparable in magnitude to typical particle-physics scales,
rare examples will have much smaller densities. Since large
values of the vacuum density will inhibit structure
formation, observers will tend to occur in models where the
vacuum density falls in a small range about zero -- thus
potentially solving both the scale and why-now problems. 
The basic form of this argument as an
upper limit on the magnitude of $\Lambda$ has existed for a
number of years: see e.g. Linde (1984; 1987) or Barrow \& Tipler (1986).
A significant further step was taken by
Weinberg (1989), who used the anthropic argument
to make the impressively bold prediction that $\Lambda$ would
indeed turn out to be non-zero at about the observed level.
Weinberg's reasoning was taken up in more
detail by Efstathiou (1995) and by
Martel, Shapiro \& Weinberg (1998).
Efstathiou calculated the expected
distribution for $\Omega_v$ for a typical observer
(a term whose meaning is discussed below); he found
a result that peaked around $\Omega_v\simeq 0.9$, in which the
observed $\Omega_v=0.75$ would not be surprising.

This is an impressive result, but it has two points at
which further study is merited.
Although he gave the general argument for the suppression of
large observed values of $\Lambda$,
Efstathiou then fixed the CMB temperature at its observed
value in order to calculate a probability distribution for
the observed value of $\Omega_v$.
In principle, it is possible that $T=2.725$~K is not a typical
value when all observers are considered, and we want to be sure that
we have not brought the why-now problem in again by the back door.
Garriga, Livio \& Vilenkin (2000) have argued that anthropic
selection does indeed provide a general solution to the why-now problem,
but it seems useful to make this argument explicit in terms of observables.
One aim of this paper is therefore to revisit Efstathiou's
calculation to calculate the joint distribution of
$\Lambda$ and $T$, to see how typical our observed location
on this plane may be. We also take the chance to apply more
accurate approximations for gravitational collapse of cosmic
structures than the usual Press-Schechter (1974) approach,
which underestimates the abundance of the most
massive objects by a factor 10.

A larger issue is the behaviour for
negative $\Lambda$, which is examined in some detail. Efstathiou
concentrated on models that were in the expanding phase, whereas
a universe with negative $\Lambda$ will eventually cease expanding and
collapse to a big crunch. 
New accurate approximations are given for the growth of density
fluctuations, for either sign of $\Lambda$.
It turns out that structure formation
in the collapsing phase is highly efficient, which presents
something of a puzzle given the observed positivity of $\Lambda$.

Finally, the same anthropic approach can in principle also be
applied to the case of cosmological curvature. We indicate how this
argument might have been applied to explain an open universe in the
period before evidence for $\Lambda$ emerged.
The paper concludes by discussing the testability of anthropic reasoning
in the light of these results.

\sec{ANTHROPIC SELECTION AND GRAVITATIONAL COLLAPSE}

\ssec{Principles}

The first step in dealing with ensembles of universes is to decide
what will be allowed to vary. In principle, nothing is guaranteed
to be fixed, but it makes sense to start in as restrictive a way
as possible. We therefore follow Efstathiou (1995): he assumed
that all members of the ensemble are exactly spatially flat,
and that the key dimensionless ratios of cosmology are as
we observe. These are (1) the photon to baryon number density
ratio; (2) the dark matter to baryon density ratio; (3) the
horizon-scale amplitude of density fluctuations, $\delta_{\ss H}$. The only variation
to be considered is in the value of the effective cosmological constant.

Many authors have allowed a wider set of parameters to vary. For
example, Garriga \& Vilenkin (2006) consider joint variations
in $\Lambda$ and in $\delta_{\ss H}$. The most radical
view is that of the string-theory landscape, in which all of physics
is free to vary (Susskind 2003), and aspects of this picture have
been explored by e.g. Tegmark et al. (2006) and Graesser \& Salem (2006).
One approach to this question is an experimental one: if the
simplest forms of anthropic variation can be ruled out, this
might be taken as evidence in favour of the landscape picture.
We therefore consider only the simplest picture here. This
amounts in practice to considering the observed universe at some high
temperature, so that any vacuum density of practical interest
is negligible, and then considering the future evolution of
copies of this universe in which the vacuum density is set to different
levels.

The most difficult and controversial question with any
ensemble of universes is how to set a measure: what is the relative
weighting of the different members of the ensemble? In the
present context, we need to know the prior probability to
assign to a given value of $\Lambda$, which will be modified
by an `observer-bias' factor, reflecting the relative
difficulty of forming observers as a function of $\Lambda$.
Various proposals for the
measure in the case of eternal inflation have been considered
by Aguirre, Gratton \& Johnson (2006). In the end,
we are persuaded by the original view of Weinberg (1989),
who took the prior on $\Lambda$ to be uniform
over a small range around zero (see also
Efstathiou 1995 and Weinberg 2000).
If there is nothing special about a value
zero, this seems a defensible assumption
given that the range of interesting densities
is tiny by comparison with particle-physics scales.

These different universes are assumed to receive a weight
according to the number of observers that exist in
them, so that cosmologists asking questions here and now are treated
as randomly selected from the totality of observers
over all universes and all times. The
exact meaning of this intuitive idea of a `typical' observer
is not so easily made precise, however, since observers are not standard
objects (what relative weight do we assign to a modern human, a caveman
and a cat?). Some of these difficulties are reviewed comprehensively
by Neal (2006). In the present instance, we can side-step many
of these issues by exploiting the assumed similarity of the members
of the ensemble in their non-vacuum physics. We do not need to
predict the absolute number of observers, nor how they are divided into
different types of observer:
it is sufficient to assume that a model with twice
as many stars is twice as likely to be experienced.
Thus, we take the weighting of each member of the ensemble to be given by
the fraction of the baryons that are incorporated
into nonlinear structures.

Weighting by collapse fraction is a
common assumption in work of this sort, although not universally
accepted. The most recent dissenters are Starkman \& Trotta (2006),
who calculate the maximum number of observations a given observer can
ever make, which introduces a further dependence on $\Lambda$.
There is certainly an ambiguity worth debating here: given two universes
that make the same number of observers, but where one set live much
longer than the other, should we give an equal weight to each, or
weight them in proportion to lifetime? Starkman \& Trotta effectively
treat equal time intervals as equally probable, but the instants
in the life of a given individual are not independent: any of us
will always answer ``yes'' to the question, ``is $\Lambda$ surprisingly
low?'', and it should not matter how many times we are able to ask.
Therefore, we prefer to weight by the numbers of observers produced,
and it seems reasonable to tie this to the number of stars that are
formed.

As discussed earlier, we need a prior for the different values of the
vacuum density: following
Efstathiou (1995) and Weinberg (1989; 2000), this is taken as being uniform
over a range around zero. We therefore
weight universes with an overall posterior probability
$$
dP(\rho_v) \propto f_c\; d\rho_v,
$$
where $f_c$ is the collapse fraction: the proportion of mass
in the universe that has become incorporated into sufficiently
large nonlinear objects.

In more detail, we will be interested in the weight attaching to
different times in each universe. This is in part a simple issue,
since we can calculate what fraction of the mass undergoes collapse
in any given time interval, so that
$$
dP(\rho_v,t) \propto {d f_c\over dt}\; dt \; d\rho_v.
$$
But this gives the time distribution for the {\it formation\/} of sites at
which life might subsequently emerge.
The more serious challenge lies in predicting the
history of observers following a formation event: how
many observers will eventually result, and what will their
distribution in time be? Rather than making arbitrary
assumptions, we can avoid the worst uncertainties by
turning the problem backwards. We can calculate
the distribution of times at which stars form in the universe,
and we know when the star with which we are associated was formed:
4.6~Gyr ago. That time corresponds to a redshift 0.457,
at which point the cosmological parameters were
$T=3.97$~K and $\Omega_v=0.49$.
We can therefore concentrate on the more concrete question of whether
the sun formed at a typical point in comparison to all stars in the multiverse.
This presumes that the subsequent history of life around each star
has no dependence on $\Lambda$, but further biological assumptions
are not required.

\japfig{1}{1}{fig1.ps}
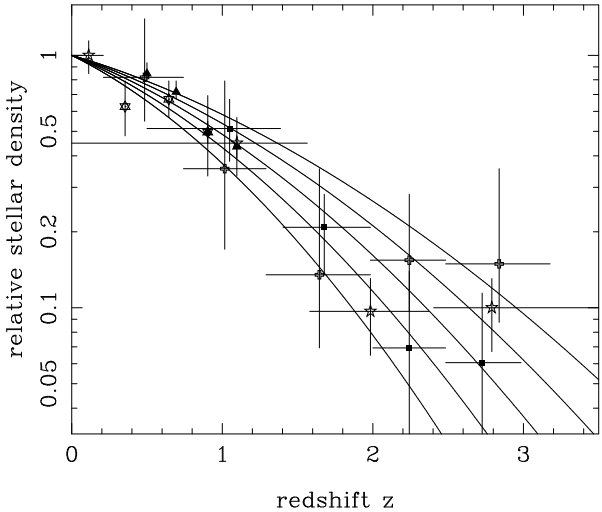
{Time dependence of star formation predicted in a simple collapse model.
The total stellar density produced by a
given epoch is assumed to scale with the total collapse fraction
associated with a single mass scale.
The density-fluctuation parameter $\nu(T=1000)$ is varied by up to 20 per
cent either side of its canonical value $\nu=250$. This
yields a good match to the data on the empirical redshift dependence of the
total stellar mass density, taken from Merloni, Rudnick \& Di Matteo (2004).}

\ssec{The collapse fraction}

For this calculation, we need to be able to predict the
fraction of baryons in the universe that are processed into stars.
This is not something that can presently be calculated without
some guidance from observation. The first galaxy-sized systems to collapse
are of low mass and high density, and clearly will form some stars,
but it is expected that feedback from this initial activity will quickly
regulate the star formation in these objects. Thus in practice most
star formation is expected to occur in the largest galaxies, and
we follow Efstathiou (1995) in treating these as being defined by
a single mass scale. However, as shown below, it is now possible to deduce this
scale empirically, rather than appealing to a priori cooling arguments
such as Rees \& Ostriker (1977).

We develop below the formulae needed to calculate accurately the
fraction of the mass that has collapsed into objects of a given mass
scale or larger. As usual, the mass scale is defined by the mass
in a homogeneous universe contained within a sphere of radius $R$.
The fractional density fluctuations smoothed with such a spherical
filter have an rms value $\sigma(R)$, and the rareness of
objects of a given mass is quantified by defining
$$
\nu\equiv \delta_c/\sigma(R),
$$
where $\delta_c$ is a density threshold of order unity.
Since $\sigma$ changes with time, we need to specify an
era in order to associate $\nu$ with a given mass.
It is convenient to make the arbitrary choice of $T=1000$~K
as a reference era (matter dominates over radiation and over
any vacuum density of interest). Fig. 1 then shows that the
existing data on the evolving comoving stellar density can be
well described using a single-scale model with
$$
\nu(T=1000)=250,
$$
adopting $\delta_c=1.686$ as justified below.

This plot does not need to make any assumptions about the
matter power spectrum, but if we adopt a canonical model
($\Omega_m=0.25$, $\Omega_b=0.04$, $h=0.73$, $n_s=0.95$,
$\sigma_8=0.8$), this corresponds to an effective galaxy mass of
$$
M_g=1.9\times 10^{12}\, M_\odot.
$$
The small-scale CDM power spectrum
is rather flat:
$$
d\ln\nu/d\ln M\simeq 0.145,
$$
so the results of this calculation should
be relatively robust. Loeb (2006) is correct to point out that,
in principle, some stars could form in dwarf galaxies, but changing the
critical mass scale by a power of ten has little impact on the results.

Even so, it is clear that this simple model can be challenged, since
the typical galaxy mass arises in a complex way, where feedback
from supernovae and AGN heats gas and prevents star formation
being very efficient (the observed cumulative efficiency today
is $\Omega_*/\Omega_b \simeq 5\% - 9\%$, depending on assumptions about
the IMF: Cole et al. 2001). It is certainly possible that the operation
of such effects in objects of a given mass could depend on density and
thus on era -- so that the overall efficiency of star formation
could have a further complicated dependence on $\Lambda$. We shall
ignore this point here, but it would clearly be of interest
to investigate the issue elsewhere using detailed galaxy formation
models.

\ssec{The cosmological mass function}

Given a filtering scale corresponding to a typical galaxy, the
linear density contrast, $\sigma(R)$ can be calculated. According to
the Press-Schechter (1974) approximation, the collapse fraction
(i.e. proportion of mass contained in objects of the given mass
scale or larger) is then
$$
f_c = {\rm erfc}\left(\nu/\sqrt{2}\right);\quad \nu\equiv\delta_c/\sigma.
$$
The critical density contrast is $\delta_c\simeq 1.686$ in an Einstein--de Sitter
model. Efstathiou (1995) gives an approximate scaling as $(1-\Omega_v)^{-0.28}$,
but this is a theoretical expectation based on the spherical collapse
model. In detailed studies of numerical simulations, Jenkins et al. (2001)
found that $\delta_c$ could be treated as constant in matching theory
to the simulated mass functions. Jenkins et al. followed
Sheth \& Tormen (1999), who established that the Press-Schechter form
for the mass function was significantly in error, with too many objects
at the peak of the mass function by about a factor 1.5, and too few
at the highest masses by a power of 10. More recently,
Warren et al. (2006) have shown that the Jenkins et al. fitting
formula still contains errors of order 10\%, and they proposed a
replacement. An unsatisfactory feature of their fit is that
it predicts collapse fractions in excess of unity for small masses,
whereas one would normally prefer to assume that $f_c \rightarrow 1$
in the limit of very small masses. It is simple to cure this by
finding an analytic formula for $f_c$; this can then be
differentiated in order to find the mass function. The following
expression matches the Warren et al. fitting formula to a maximum
error of about 1\% over the whole range where data exist:
$$
f_c=(1+a\,\nu^b)^{-1}\exp(-c\,\nu^2),
$$
where $(a,b,c)=(1.529,0.704,0.412)$.

\ssec{Growth of density perturbations}

For this exercise, we also require the linear growth function
for density perturbations, which
can be expressed as a numerical integral (Heath 1977). It is convenient to
have an accurate numerical approximation, and the following expressions
are good to a maximum error of 0.1\%. The cases of positive and
negative $\Lambda$ are somewhat distinct. For the positive case,
$$
\delta(a) \simeq x(1-x^{1.91})^{0.82} + 1.437\left(1-(1-x^3)^{2/3}\right),
$$
where $x$ denotes $\Omega_v(a)^{1/3}$, and we choose the $a=1$ point
to correspond to equal density in matter and vacuum:
$$
\Omega_v(a) = (1+a^{-3})^{-1},
$$
so that $\delta(a) \simeq a$ for small $a$. For a starting point where
$\Omega_v(a)$ is small, the total amount of growth is
$$
\delta_\infty/\delta_{\rm init} \simeq 1.437/a_{\rm init}
\simeq 1.437/[\Omega_v(a_{\rm init})]^{1/3}.
$$
Since $\Omega_v(a_{\rm init}) = \rho_v / 5.375\times10^8\, {\rm meV}^4$ for
our choice of $T=1000$~K as a normalization point,
at which the galaxy-scale fluctuation is $\nu=250$, this immediately
allows the asymptotic value of $\nu$ to be deduced.

For the negative density case, we need time as a coordinate, since the
scale factor is not monotonic:
$$
a(t) = \left[\sin(3t/2)\right]^{2/3},
$$
where here we choose units such that $a=1$ at the point of maximum
expansion, and time is measured in units of $(8\pi G|\rho_v|/3)^{-1/2}$,
so that Friedmann's equation is $(\dot a/a)^2 = a^{-3} -1$
and $\Omega_v(a) = (1-a^{-3})^{-1}$.
Here, the approximation for the growth function is
$$
\delta(t) ={ (3t/2)^{2/3} \over
(1+0.37(t/t_{\rm coll})^{2.18})\, (1-(t/t_{\rm coll})^2) }.
$$
Again, the normalization
is that $\delta(a)\simeq a$ for small $a$.
Note that the fluctuations diverge at the collapse time
($t_{\rm coll}=2\pi/3$) as $1/(t_{\rm coll}-t)$: this corresponds
to the decaying mode in the expanding phase.

\japfig{2}{1}{fig2.ps}
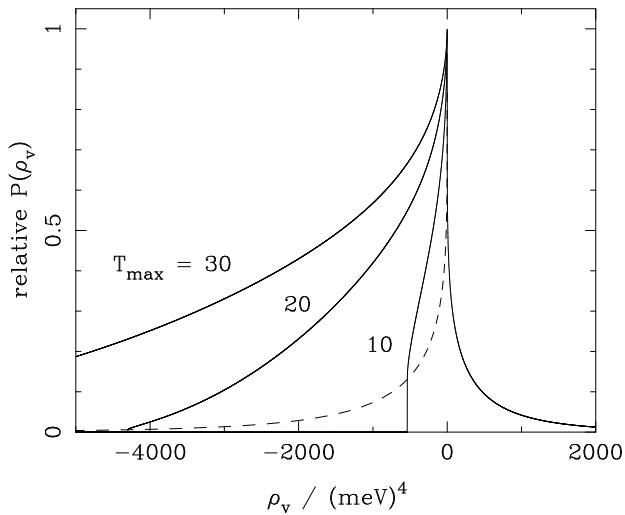
{The collapse fraction as a function of the vacuum density, which
is assumed to give the relative weighting of different models. The
dashed line for negative density corresponds to
the expanding phase only, whereas the solid lines for negative density
include the recollapse
phase, up to maximum temperatures of 10~K, 20~K, 30~K.
The observed value of the vacuum density is $33\, \rm meV^4$.
}

\sec{THE EXPECTED SIGN OF $\bf\Lambda$}

The collapsing phase turns out to be important when we consider
structure formation. A substantial negative $\Lambda$ limits the
amount of growth that can occur before the universe ceases to expand,
but the total amount of growth after this is limited only by
how close to the big crunch we are prepared to venture. Eventually,
$f_c$ tends to unity in all such recollapsing models.
Since the uniform prior extends arbitrarily far towards more negative
values of $\Lambda$, this apparently implies that all the weight should
be given to models with $\Lambda<0$.

However, structures that form very close to the final
singularity are not of interest for the anthropic
calculation: there is little time remaining for life to
develop, and in any case the CMB will have heated up to the
point where it interferes with life -- or indeed perhaps
even with the formation of stars and planets themselves.
It is simplest to express this cutoff in the recollapsing
phase in terms of a maximum temperature that we are willing
to consider, although this can be directly translated to a limit
on the time remaining before the big crunch. Since the
recollapsing phase is the time-reversed version of the
expansion, the time remaining from temperature $T$
until the big crunch is just what would have elapsed
from the big bang until this temperature. Normally, the
matter-dominated approximation will apply, so
$$
t(T)\simeq {2\over 3 H_0}\, \Omega_m^{-1/2}\, (1+z)^{-3/2}
= \left({T\over 18.6\;\rm K}\right)^{-3/2} \; {\rm Gyr}.
$$
We know from observations that star formation in galaxies can proceed
actively at redshift $z\simeq 7$, so $T_{\rm max}>10$~K
on these grounds. This would leave only a few Gyr
after formation for life to evolve, so presumably $T_{\rm max}$
should not be much larger than this, and could well be smaller.
This is not so much a biological argument as one based on
stellar lifetimes. Highly negative values of $\Lambda$ would
cause the universe to turn round without ever cooling below
this critical temperature, so these models may be excluded from
the point of view of generation of observers, even though they
form galaxy-scale structures efficiently. This is completely distinct from
the situation at highly positive $\Lambda$, where the problem is
the failure to create structure at any time.

Rather than singling out a particular value of $T_{\rm max}$,
we may as well perform the calculation allowing
it to take a range of values.
This is illustrated in Fig. 2, which shows the relative weight to be
given to models as a function of $\rho_v$. We can integrate this
distribution to obtain
Fig. 3, which shows  how the probability of observing
a negative $\Lambda$ varies with the maximum temperature for
structure formation that we are willing to tolerate.
The probability that observers inhabit a universe with
$\Lambda<0$ is about 50\% if $T_{\rm max}=8.5$~K,
or as little as 1\% if $T_{\rm max}=1.5$~K.
Conversely, only about 6\% of observers will experience a
{\it positive\/} $\Lambda$ if $T_{\rm max}=30$~K; but this
is probably too tolerant of late-forming observers.
From Fig. 2, it is furthermore clear that the
bulk of any weight in favour of negative $\Lambda$ lies with
the recollapsing phase: the probability of inhabiting the
expanding phase is approximately the same as the probability
that $\Lambda$ is positive.
According to the anthropic framework, it is therefore far from inevitable
that we have ended up in a positive-vacuum expanding
universe, although neither is it particularly unusual.

\japfig{3}{1}{fig3.ps}
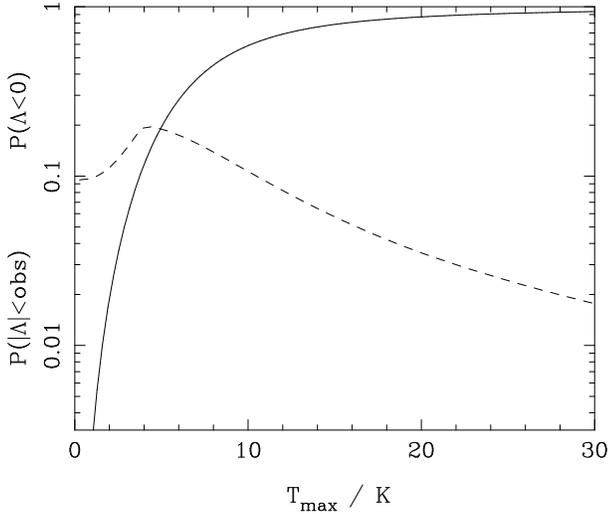
{Integrating under the distribution of Fig. 2, we can deduce the
probability of inhabiting a universe with $\Lambda<0$, as a function
of the assumed maximum temperature for galaxy formation (solid line).
The dashed line shows, under the same assumption, the probability
that the absolute value of $\Lambda$ would lie within its observed
value of $\rm (2.39\, meV)^4$. For a maximum temperature of 8.5~K, positive
and negative values of $\Lambda$ are equally probable, but negative
$\Lambda$ is disfavoured by more stringent limits on temperature: for
a maximum temperature of 1.5~K, only about 1\% of observers will inhabit
a universe with $\Lambda<0$.}

We can now make a first attempt to assess how well this
anthropic prediction matches reality. Since $\Lambda=0$ is a
special point, it is reasonable to consider the frequentist question:
`what is the probability that $|\rho_v|$ lies within
$(2.39\,{\rm meV})^4$ of the origin?'. The answer is also plotted
in Fig. 3: about 10\% if we ignore negative $\Lambda$, peaking at
20\% for $T_{\rm max}=4$~K, and declining for larger values.
The consistency of observations
with anthropic prediction therefore depends somewhat
on the recollapsing phase. In what follows, we will concentrate
on the expanding case of positive $\Lambda$, but it
should be borne in mind that any anthropic probabilities we
deduce subsequently should be reduced slightly to account
for the fact that observers also have a non-negligible chance of being found
in a recollapsing model.

\japfig{4}{1}{fig4.ps}
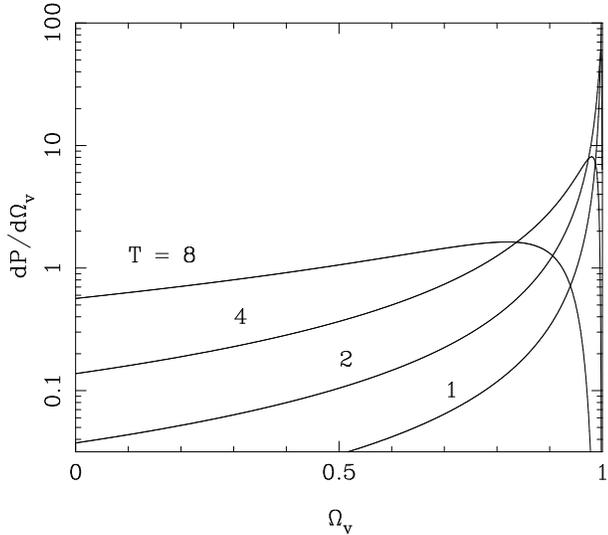
{The probability distribution of $\Omega_v$ at observed temperatures
of  1~K, 2~K, 4~K, 8~K, with higher temperatures pushing the distribution
to lower values of $\Omega_v$. This plot should be contrasted with
Fig. 2 of Efstathiou (1995).}

\sec{THE JOINT DISTRIBUTION OF $\bf\Lambda$ AND T}

The above calculation addresses the scale problem, but only considers
the total amount of structure formation, not when it occurs.
This problem was considered by Garriga, Livio \& Vilenkin (2000),
who calculated the posterior distribution on the plane
$(t_\Lambda, t_{\ss G})$, where $t_\Lambda$ is the time of 
$\Lambda$-domination and $t_{\ss G}$ the time at which a typical galaxy was formed.
They used a Press-Schechter approach to show that the probability density on this plane peaks
around $t_{\ss G}\sim t_\Lambda$, so that the why-now problem is solved.
We need something similar, but we want to see explicitly how the
Efstathiou analysis is affected by a given assumed temperature; it is therefore
necessary to cast the distribution in terms of the observables $\Omega_v$ and $T$.
We have the differential probability distribution
$dP \propto d\rho_v\, df_c$, and we want to change variables to
$\Omega_v$ and $T$. Rather than doing this in a single step by working
out the Jacobian of the transformation, we can first note that
the collapse fraction is just a function of $T$ for given $\rho_v$, so that
$$
dP \propto d\rho_v\, \left.{\partial f_c\over \partial T}\right|_{\rho_v}\, dT.
$$
If we now change from ($\rho_v$,$T$) to ($\Omega_v$,$T$), the Jacobian is diagonal,
and
$$
dP \propto \left.{\partial \rho_v\over \partial \Omega_v}\right|_{T}\,
\left.{\partial f_c\over \partial T}\right|_{\rho_v}\; d\Omega_v\; dT.
$$
To evaluate the first factor on the rhs, note that
the vacuum density parameter at some early time $t_0$ is
$$
\Omega_{v0} = {\Omega_v \over
\Omega_v + (1-\Omega_v)a_0^{-3} }
\simeq {\Omega_v \over
(1-\Omega_v)a_0^{-3} },
$$
where $\Omega_v$ is the density parameter at the later time of interest,
and $a_0 = T/T_0 \rightarrow\infty$ gives the latter approximation.
In this limit, $\Omega_{v0} \ll1$; this is therefore proportional to $\rho_v$, giving
$$
\left.{\partial \rho_v\over \partial \Omega_v}\right|_{T} \propto { T^3 \over (1-\Omega_v)^2}.
$$
This gives higher weight to $\Omega_v$ close to 1, since this is
an attractor for the evolution of $\Omega_v(t)$ if $\Lambda>0$.
The focusing towards $\Omega_v=1$ increases as $T$ falls, and this is
reflected in the $T^3$ factor. As a result, $dP/dT$ with $\Omega_v$
fixed at zero differs in this framework from what we would
normally calculate for an Einstein--de Sitter universe.

Efstathiou (1995) did not need to consider the $T^3$ factor,
since he held the temperature constant at its observed value.
This is appropriate if we want to use the anthropic framework
in a Bayesian sense, to make the best {\it prediction\/} of the current
value of $\Omega_v$ given what else we know. For similar reasons,
Efstathiou held constant the observed large-angle CMB fluctuations,
whereas these vary with temperature (i.e. with time of observation).
Here, we are interested in the broader question of whether the
conditions we observe are close to those experienced by a typical observer.
To show that this distinction matters, consider Fig. 4. This plots the
posterior probability of $\Omega_v$ for various choices of the
observed temperature, and shows that the result is sensitive
to temperature. For $T\simeq 8$~K, the distribution peaks
near the observed $\Omega_v=0.75$, but for lower $T$ the distribution
is dominated by the spike in the prior at $\Omega_v=1$.

\japfig{5}{1}{fig5.ps}
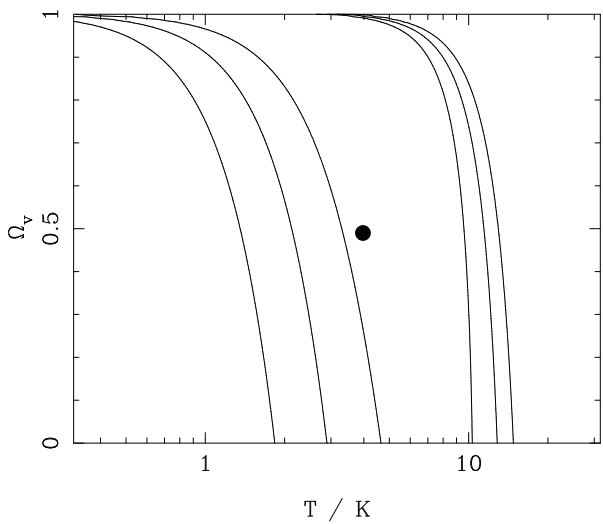
{Contours of probability density on the $\log(T)-\Omega_v$ plane.
The three contours shown enclose 68\%, 95\% and 99\% of the probability.
The solid point shows the conditions at the epoch of formation of the sun.}

We now return to the full joint distribution for $\Omega_v$ and $T$,
which we had in the form
$$
dP \propto { T^3 \over (1-\Omega_v)^2}\,
\left.{\partial f_c\over \partial T}\right|_{\rho_v}\; d\Omega_v\; dT.
$$
The remaining partial derivative is
$$
\left.{\partial f_c\over \partial T}\right|_{\rho_v} =
{\partial f_c\over \partial \ln\nu}\, T^{-1} \, {\partial \ln \nu \over \partial \ln T}.
$$
For convenience, we can use the Peebles (1980) approximation for
the logarithmic growth rate:
$$
{\partial \ln \nu \over \partial \ln T} \simeq (1-\Omega_v)^{0.6},
$$
so that overall
$$
dP \propto {\partial f_c\over \partial \ln\nu} \, T^2\, (1-\Omega_v)^{-1.4}
\; d\Omega_v\; dT.
$$

This joint probability distribution is show in Fig. 5, converting
to $\log T$ for convenience. The `observed' universe of
$(T, \Omega_v)=(3.97,0.49)$ is plotted as a point. It is
clear that the point does not lie in a particularly unusual
position in this plane. If we draw contours of constant
probability density, the point lies well within the
68\% contour. These contours are not unambiguous, as they
depend on the measure adopted on the $\Omega_v - T$ plane.
However, if we inspect the marginalized distributions
for $\Omega_v$ and $T$,
shown in Figs 6 \& 7, we see that the observed conditions
are close to the 50\% point in each quantity. In short, the
anthropic calculation suggests that we are indeed extremely
typical observers, both in terms of the vacuum density we
see, and when we see it.

\japfig{6}{1}{fig6.ps}
{The integral probability distribution for $\Omega_v$,
marginalized over $T$.}

\japfig{7}{1}{fig7.ps}
{The integral probability distribution for $T$,
marginalized over $\Omega_v$.}

\sec{ANTHROPIC ARGUMENTS APPLIED TO CURVATURE}

So far, we have assumed that the vacuum is indistinguishable from a cosmological
constant, with equation of state $w=P/\rho c^2=-1$. We know that this
is a good approximation for our universe, and it
would in any case go beyond the
scope of this paper to consider ensembles in which more than one
parameter varies. However, it is worth paying some attention to the
special case $w=-1/3$. With the critical exception of the distance-redshift
relation, a flat model with this vacuum equation of state is indistinguishable
from a model with non-zero spatial curvature: the Friedmann equation and
the growth equation for density perturbations are identical. We can then
use a modified version of the above approach to show what happens if we
confront anthropic reasoning with the curvature of the universe.

To some extent, curvature presents a parallel set of problems
to the vacuum. There is a scale problem, in the sense that natural
initial conditions might be thought to have a total
$|\Omega-1|$ of order unity, which would lead to a universe dominated by
curvature long before today. There could also be a why-now problem, if
the present curvature was non-zero at the level
of $|\Omega-1|\sim 0.01$, which cannot
currently be excluded. It is commonly assumed that inflation
solves the curvature scale problem and also predicts that there
is no why-now problem, but it is interesting to see how the
anthropic apparatus copes with this case.
This issue has been examined previously
(e.g. by Garriga, Tanaka \& Vilenkin 1999), but it is of some interest
to see how our specific approach works out in this case.
Rather than break our rule of allowing only one parameter to
vary in the ensemble, we take a historical approach and
imagine how anthropic arguments might have been applied to curvature
decades ago, when many cosmologists were convinced that $\Lambda=0$
(it is perhaps surprising that anthropic ideas received little emphasis at this time).

Unlike the vacuum density, curvature lacks an obvious time-independent
absolute scale. At any given era, one can define
$\Omega_k\equiv 1-\Omega_m$, so it will be convenient to consider
$\Omega_k(T=1000)$ as our parameter. This evolves as
$\Omega_k(a)=\Omega_k/(\Omega_k+\Omega_m/a)$.
Interesting values of this number at the reference $T=1000\,\rm K$
will be small, so it is tempting to follow our previous
procedure and assign a uniform
prior around zero, and weight models by their asymptotic collapse factor.
In the days before inflation, this might have been a defensible
expression of ignorance: the essence of the flatness problem is
that order unity positive or negative curvature in the initial conditions
seems more natural than the tiny amount necessary to yield an
almost flat universe today. But in modern models where `pocket' universes
are formed by tunnelling, the result is an open universe, so
priors on curvature might well have a discontinuity at zero
(e.g. Freivogel et al. 2006). The idea of a uniform prior for
curvature is therefore less well founded than it is for $\Lambda$.
Nevertheless, it is of some interest to carry out the exercise
of adopting a uniform prior and seeing where it leads.

For models with $\Lambda=0$, the perturbation growth as a function of $a$ is analytic:
$$
\eqalign{
\delta(a) &=
+1 + {3\over a^{3/2}}\left(\sqrt{a}-\sqrt{1+a}\,{\rm sinh}^{-1}\sqrt{a}\right)\quad (\Omega_k>0) \cr
&=
-1 + {3\over a^{3/2}}\left(\sqrt{a}-\sqrt{1-a}\,{\rm sin}^{-1}\sqrt{a}\right)\quad\;\; (\Omega_k<0); \cr
}$$
in the collapsing phase for negative $\Omega_k$, ${\rm sin}^{-1}\sqrt{a}$
is replaced by ${\rm sin}^{-1}\sqrt{a}-\pi$. These expressions are normalized
so that $\delta \simeq 2a/5$ for $a\ll1$. The convention for the scale factor assumes
that $a$ can be written in terms of conformal time, $\eta$ as
$$
\eqalign{
a &= (1-\cos\eta)/2  \quad {\rm (closed)} \cr
a &= (\cosh\eta-1)/2  \quad {\rm (open)}, \cr
}
$$
so that $a=1$ at maximum expansion in the closed model. The relation to
density parameters in these units is
$$
\eqalign{
\Omega_k(a) &= a/(a-1)  \quad {\rm (closed)} \cr
\Omega_k(a) &= a/(a+1)  \quad {\rm (open)}. \cr
}
$$

We can now repeat the exercise of Fig. 2 for the case of curved universes
with $\Lambda=0$, and the results are shown in Fig. 8. Generally, the
anthropic weighting as a function of curvature looks similar to the
weighting as a function of $\Lambda$, but with some important differences.
The asymmetry in favour of recollapsing models is not so extreme:
the probability of experiencing $\Omega_k >0$ (i.e. a negatively curved
open universe) is 41\% for $T_{\rm max}=10$~K, falling to 29\%
for $T_{\rm max}=30$~K.

What about the magnitude of curvature?
A decade ago, open models were seriously under consideration,
and some would have argued for $\Omega_k \simeq 0.7$, so that
$\Omega_k(T=1000) \simeq 0.006$. From Fig. 8, we see that the
typical curvature predicted for such $\Lambda$-free universes
was $\Omega_k(T=1000) \simeq 0.01$, so an anthropic approach to explaining
the density parameter in matter-only models would have yielded
sensible answers. In order to reject an anthropic explanation for an
open universe at the 1\% level, it would have been necessary to have a limit
of $|\Omega_k(T=1000)| \ls 10^{-4}$
(depending slightly on limiting temperature), corresponding to
$|\Omega_k| \ls 0.035$ at $T=2.725$.
In fact, we barely know that the universe is flat to this
precision even today: the present limit is
approximately  $|\Omega_k|<0.02$, according to Spergel at al. (2006).
Therefore, an anthropic approach to curvature would have been
perfectly consistent with 1990s data.

Today, the issue of curvature would be approached in the context
of inflation, where a sufficient number of $e$-foldings, $N$, of the
expansion will lead to values of the present
curvature that are unmeasurably small. 
The issue of interest is therefore the prior to be placed
on $N$ (see e.g. Freivogel et al. 2006).
Whatever the result of such a calculation, however, a
sufficiently small upper limit on curvature would 
allow the anthropic argument to be rejected.
Similarly, if we had no detection of $\Lambda$, our
earlier results show that a sufficiently strong upper limit
would reject the anthropic approach, leading us to require
a physical mechanism that forces $\Lambda=0$. Anthropic
reasoning is thus testable and could point to new physics.
But this is not the situation we face:
we have an actual detection of $\Lambda$,
rather than an ever-retreating upper
limit, and no a priori theory predicts the observed number.
An explanation in terms of anthropic selection from an
ensemble matches what we see, and so far there is no credible alternative.

\japfig{8}{1}{fig8.ps}
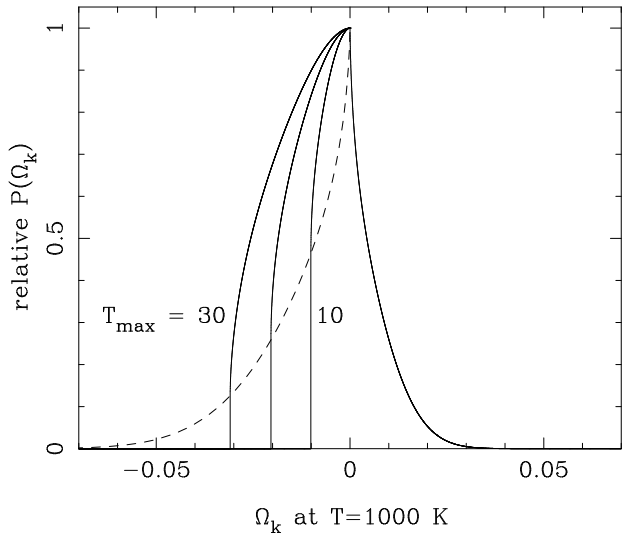
{The collapse fraction as a function of the curvature
in models with $\Lambda=0$, which
is assumed to give the relative anthropic weighting of different models.
Curvature is specified as $\Omega_k=1-\Omega_m$ at the era when $T=1000$~K.
The dashed line for negative $\Omega_k$ corresponds to
the expanding phase only, whereas the solid lines for negative $\Omega_k$
include the recollapse
phase, up to maximum temperatures of 10~K, 20~K, 30~K.}

\sec{CONCLUSIONS}

This paper has revisited the anthropic calculation of Efstathiou (1995)
in more detail, dropping the assumption that the CMB temperature is
fixed, and considering also the formation of structure during the
collapsing phase of a model with negative vacuum density.
We adopt Weinberg's (1989) assumption that the prior on $\Lambda$
is uniform in a small range around zero, so that models are
weighted simply by their collapse function.
We have attempted to avoid issues such as the lifespan of civilizations by
using the formation era of the sun as the data to explain: so long as observers
eventually form with some fixed mean number per star, we can
ignore how they are subsequently distributed in time.
For universes that are expanding, the conclusion
is that the sun formed at completely typical values of $\Omega_v$ and
the CMB temperature.
There is therefore no observational reason to challenge the idea of a
uniform prior on $\Lambda$, nor to require variation of other
parameters within the ensemble -- although there remains
theoretical motivation for considering more complex alternatives.

This approach can certainly be questioned in the case of negative
vacuum densities, where the age of the universe is finite. One might
be tempted to argue that all the weight should go to positive-density
universes, since civilizations can then potentially last forever.
However, this is too optimistic, since the event horizon in models
that are asymptotically de Sitter limits the resources that are available.
Once most of the stars in a given model have formed, it is reasonable
to expect that observers will find existence progressively more difficult
after a further few billion years as the existing stars die out.
In any case, we know that we are members of a civilization that has not
yet outlived its star; the question of whether some
races might live for a trillion years can therefore
be ignored. The reasoning here is the same as in the ``God's coin toss''
thought experiment discussed by Olum (2002). This experiment imagines
that according to the toss of a coin either 10 (heads) or 1000 (tails)
people are created
and given a number. If you have no knowledge of your number, your odds
for the result of the coin toss should be 100:1
in favour of tails; but if your number is $\le10$,
your odds should be equal.

From this point of view, there thus
seems no reason not to give appropriate weight to the
flourishing of observers in recollapsing models.
In contrast to models with positive vacuum density,
gravitational collapse is always perfectly efficient
in models where $\Lambda<0$. But this does not imply that
anthropic selection must favour this case:
temperatures should not too be too high at putative formation,
and there should remain at least a few billion years before
further collapse renders the CMB
hot enough to be an environmental hazard.
These criteria motivate a maximum temperature of order 10~K,
for which only a minority ($\sim 10\%$)
of observers should find themselves in recollapsing models.
This was not our fate; but it is interesting to speculate how
observational cosmology might have developed in such a case.

\section*{ACKNOWLEDGEMENTS}

This project was carried out under the
support of a PPARC Senior Research Fellowship.
Thanks are due to Bernard Carr, Andrei Linde, Roberto Trotta
Alex Vilenkin and particularly George Efstathiou for comments
in the course of this work.

{

\pretolerance 10000

\section*{References}

\japref Aguirre A., Gratton S., Johnson M.C., 2006, hep-th/0611221
\japref Barrow J.D., Tipler F.J., 1986, {\it The Anthropic Cosmological Principle\/}, Oxford University Press
\japref Carter B., 1974, Proc. IAU Symposium no. 63 (Reidel), p291
\japref Cole S. et al. (the 2dFGRS Team), 2001, MNRAS, 326, 255
\japref Efstathiou G., 1995, MNRAS, 274, L73
\japref Freivogel B., Kleban M., Rodriguez Martinez M., Susskind L., 2006, JHEP 0603, 039 (hep-th/0505232)
\japref Garriga J., Tanaka T., Vilenkin A., 1999, Phys. Rev. D, 60, 023501
\japref Garriga J., Livio M., Vilenkin A., 2000, Phys. Rev. D, 61, 023503
\japref Garriga J., Vilenkin A., 2000, Phys. Rev. D, 61, 083502
\japref Garriga J., Vilenkin A., 2006, Prog. Theor. Phys. Suppl. 163, 245 (hep-th/0508005)
\japref Graesser M.L., Salem M.P., 2006, astro-ph/0611694
\japref Heath D., 1977, MNRAS, 179, 351
\japref Jenkins A.R., Frenk C.S., White S.D.M., Colberg J.M. Cole S., Evrard A.E., Couchman H.M.P., Yoshida N., 2001, MNRAS, 321, 372
\japref Liddle A.R., Scherrer R.J., 1999, Phys. Rev. D 59, 023509
\japref Linde A.D., 1984, Rept. Prog. Phys., 47,  925
\japref Linde A.D., 1987, in {\it 300 Years of Gravitation}, ed. S.W. Hawking and W. Israel, Cambridge University Press, p604
\japref Loeb A., 2006, JCAP, 0605, 009 (astro-ph/0604242)
\japref Malquarti M., Copeland E.J., Liddle A.R., 2003, Phys. Rev. D, 68, 023512
\japref Martel H., Shapiro P., Weinberg S., 1998, ApJ, 492, 29
\japref Merloni A., Rudnick G., De Matteo T., 2004, MNRAS, 354, L37
\japref Neal R.M., 2006, math.ST/0608592
\japref Olum K.D., 2002, Phil. Q., 52, 164 (gr-qc/0009081)
\japref Peacock J.A., 1999, {\it Cosmological Physics\/} (Cambridge)
\japref Peebles P.J.E., 1980, {\it The large-scale structure of the universe\/} (Princeton)
\japref Press W.H., Schechter P., 1974, ApJ, 187, 425
\japref Rees M.J., Ostriker J.P., 1977, MNRAS, 179, 541
\japref Sheth R.K., Tormen G., 1999, MNRAS, 308, 119
\japref Spergel D.N. et al., 2006, astro-ph/0603449
\japref Starkman G.D., Trotta R., 2006, astro-ph/0607227
\japref Susskind L., 2003, hep-th/0302219
\japref Tegmark M., Aguirre A., Rees M.J., Wilczek F., 2006, Phys.Rev. D73, 023505 (astro-ph/0511774)
\japref Warren M.S., Abazajian K., Holz D.E., Teodoro L, 2006, ApJ, 646, 881 (astro-ph/0506395)
\japref Weinberg S., 1989, Rev. Mod. Phys. 61, 1
\japref Weinberg S., 2000, Phys. Rev. D, 61, 103505

}

\bye